\documentclass[11pt,superscriptaddress,aps,prd,preprint]{revtex4}
\usepackage{amsmath}

\makeatletter
\usepackage[T1]{fontenc}
\usepackage{amsmath}
\usepackage{amssymb}
\usepackage{graphicx}

\usepackage{color}

\usepackage{slashed}

\DeclareSymbolFont{extraitalic}      {U}{zavm}{m}{it}
\DeclareMathSymbol{\Qoppa}{\mathord}{extraitalic}{161}
\DeclareMathSymbol{\qoppa}{\mathord}{extraitalic}{162}
\DeclareMathSymbol{\Stigma}{\mathord}{extraitalic}{167}
\DeclareMathSymbol{\Sampi}{\mathord}{extraitalic}{165}
\DeclareMathSymbol{\sampi}{\mathord}{extraitalic}{166}
\DeclareMathSymbol{\stigma}{\mathord}{extraitalic}{168}

\newcommand{\bea}{\begin{eqnarray}}
\newcommand{\eea}{\end{eqnarray}}

\newcommand{\pr}[1]{\ensuremath{\left[#1\right]}}
\newcommand{\pc}[1]{\ensuremath{\left(#1\right)}}

\newcommand{\R}{\mathcal{R}}
\newcommand{\G}{\tilde{\Gamma}}

\begin{document}

\title{G\"{o}del-type solutions in hybrid metric-Palatini gravity }

\author{J. S. Gon\c{c}alves}\email[]{junior@fisica.ufmt.br }
\affiliation{Instituto de F\'{\i}sica, Universidade Federal de Mato Grosso,\\
78060-900, Cuiab\'{a}, Mato Grosso, Brazil}

\author{A. F. Santos}\email[]{alesandroferreira@fisica.ufmt.br}
\affiliation{Instituto de F\'{\i}sica, Universidade Federal de Mato Grosso,\\
78060-900, Cuiab\'{a}, Mato Grosso, Brazil}

\begin{abstract}
In this paper, the hybrid metric-Palatini gravity is an approach to modified gravity in which is added to the usual Einstein-Hilbert action a supplementary term containing a Palatini-type correction of the form $f(\R,T)$. Here, $\R$ is the Palatini curvature scalar, which is constructed from an independent connection and $T$ is the trace of the energy-momentum tensor. This theory describes a non-minimal coupling between matter and geometry. The modified Einstein field equations in this hybrid metric-Palatini approach are obtained. Then, it is investigated whether this modified theory of gravity and its field equations allow G\"{o}del-type solutions, which essentially lead to violation of causality. Considering physically well-motivated matter sources, causal and non-causal solutions are explored.
\end{abstract}

\maketitle

\section{Introduction}

Cosmological observations indicate that the universe is currently expanding at an accelerated rate \cite{Riess, Per, Adam, Cole, Anderson}. To explain the accelerated expansion of the universe, a large number of modified theories of gravity have been proposed (for a review, see \cite{Clifton}). These approaches consider more general Lagrangian functions, extra dimensions, geometric quantities, among others. These gravitational theories can be obtained from two different theoretical approaches, called the metric and the Palatini formalism. In the first formalism, the dynamic variable is just the metric tensor, while in the latter, in the Palatini variational approach, the metric tensor and the affine connection are taken as independent field variables. Both formalisms, when applied to general relativity, lead to the same gravitational field equations. However, in modified theories of gravity, this is no longer the case. For a review of modified theories of gravity in the Palatini approach, see \cite{palatini_motivation1}. Although these approaches have led to possible ways to explain the observed accelerated expansion of the universe, they do have some problems. For example, the $f(R)$ gravity has difficulties explaining the evolution of cosmological perturbations and local gravity constraints \cite{palatini_motivation1}. Then, a third approach can be considered to study the modified theories of gravity.

An alternative way to explain the current accelerated expansion of the universe is the hybrid metric-Palatini gravity \cite{first_hybrid_paper}. This theory consists of the superposition of the metric (Einstein–Hilbert action) and Palatini formalisms. Several studies have been developed with this approach, such as, classical gravitational tests in the solar system have been investigated \cite{hybrid_motivation1}, the dark energy problem has been analyzed \cite{Capo}, models of the accelerating universe without a cosmological constant have been presented \cite{Capo2}, Noether symmetries have been studied \cite{Boro}, axial gravitational perturbations and the quasi-normal modes frequencies of black holes have been studied  \cite{hybrid_motivation2}, Kerr black hole solutions and their stability have been explored \cite{Rosa}, static and spherically symmetric solutions have been discussed \cite{hybrid_motivation3, Bron, esf, esf1}, among others. In addition, pathologies such as Ostrogradsky instability have been analyzed \cite{Adria}. It is observed that the hybrid formalism does not, in general, guarantee the absence of ghosts. Furthermore, to avoid these pathologies, it is necessary to impose additional constraints on the gravity sector. For a review of hybrid metric-Palatini gravity, see \cite{Rev, Rev2, hybrid_motivation4}.

The present paper continues the study of the hybrid metric-Palatini gravity. Here, the $f(R,T)$ gravity is a\-nalyzed. This gravitational model was first proposed in \cite{first}. This model is characterized by the non-minimal coupling between matter, described by the trace of the energy-momentum tensor $T$, and geometry, defined by the Ricci scalar $R$. In the hybrid formalism, the function $f(\R,T) + R$  is introduced, where $R$ represents the Einstein–Hilbert part of the action and $\R = R(g,\G)$ is Ricci-Palatini scalar. As an application of this gravitational theory, the causality problem is investigated. Then the G\"{o}del-type solution \cite{tipo_godel1} is considered. The motivation is based on the fact that, if gravity is described by the hybrid metric-Palatini $f(\R,T)$ gravity theory, there are a number of issues that must be re-examined in its context, such as the question of whether this gravity theory allows G\"{o}del-type solutions, which are allowed in general relativity and in various modified gravity theories.

The G\"{o}del-type metric is a generalization of the G\"{o}del solution proposed by Kurt G\"{o}del in 1949 \cite{godel}. The main feature of the G\"{o}del solution is the possibility of Closed Timelike Curves (CTCs) that lead to causality violation. This causality violation does not occur locally, since in general relativity the space-times locally have the same causal structure of the special relativity. However, on a non-local scale significant differences may arise. Then, globally, a violation of causality is allowed beyond a region limited by a critical radius. These CTCs appear in other cosmological models that have hyperbolic or spherical symmetry, such as Kerr black hole, Van-Stockum model, cosmic string, among others \cite{ctc1,ctc2}. The causality problem has been verified in many gravitational theories, such as, $f(R)$ theory \cite{fr_and_godel}, $k$-essence theory \cite{kessence_and_godel}, Chern-Simons gravity \cite{chersimon_and_godel1, chersimon_and_godel2}, $f(T)$ gravity \cite{ft_and_godel}, $f(R,T)$ gravity \cite{frt_and_godel}, bumblebee gravity \cite{bumblebeee_and_godel}, Horava-Lifshitz gravity \cite{horava_and_godel}, Brans-Dicke theory \cite{brans_and_godel}, $f(R,Q)$ gravity \cite{frq_and_godel}, $f(R,\phi,X)$ gravity \cite{fr_phi_X}, $f(R,T)$ Palatini gravity \cite{frt_palatini}, among others. 

The present paper is organized as follows. In section II, an introduction to the hybrid metric-Palatini $f(\R,T)$ gravity is made. In section III, the G\"{o}del-type metric is introduced and the field equations are solved for this metric. Then, the causality problem is discussed. Different matter contents, such as perfect fluid, a combination of perfect fluid and scalar field and only a scalar field, are considered. Causal and non-causal solutions are obtained. The critical radius for non-causal solution is calculated. In section IV, remarks and conclusions are presented.

\section{Hybrid Metric-Palatini $f(\R,T)$ gravity}

In this section, the field equations for the hybrid metric-Palatini $f(\R,T)$ gravity are calculated. The action that describes this theory is given as
\begin{equation}\label{action_1}
    S=\frac{1}{2\kappa^2}\int d^4x \sqrt{-g} \pr{\ R + \ f(\R,T) + \mathcal{L}_m},
\end{equation}
where $\kappa^2 = 8\pi G$, $g$ is the determinant of the metric tensor $g_{\mu\nu}$, $R$ is the Ricci scalar associated with the metric $g_{\mu\nu}$, $f(\R,T)$ is a function explicitly dependent on the Ricci-Palatini scalar $\R = R\pc{g,\tilde{\Gamma}}$ and the trace of the energy-momentum tensor $T$ and $\mathcal{L}_m$ is the matter Lagrangian that is a function of $g$ and physical fields $\psi$. The Ricci-Palatini scalar that depends on both the metric and an independent connection $\tilde{\Gamma}$ is written as  
\begin{equation}
    \R =g^{\mu\nu} \tilde{R}_{\mu\nu}\left(\tilde{\Gamma}\right), 
\end{equation} 
where the Ricci tensor associated with the connection $\tilde{\Gamma}$ is 
\begin{equation}
    \tilde{R}_{\mu\nu}\left(\tilde{\Gamma}\right) = \partial_{\lambda} \tilde{\Gamma}_{\mu\nu}^\lambda - \partial_{\nu} \tilde{\Gamma}_{\mu\lambda}^\lambda + \tilde{\Gamma}_{\mu\nu}^\lambda \tilde{\Gamma}_{\lambda \alpha}^\alpha - \tilde{\Gamma}_{\mu\lambda}^\alpha \tilde{\Gamma}_{\nu \alpha}^\lambda.
\end{equation}

In order to obtain the field equation related to the metric field it is assumed that $\delta \tilde{R}_{\mu\nu}\pc{\tilde{\Gamma}}=0$, that is, the Palatini connection is constant, and the matter Lagrangian is independent of $\partial_\lambda g_{\mu\nu}$. Then the variation of the action (\ref{action_1}) with respect to the metric gives
\begin{equation}\label{field_1}
     G_{\mu\nu}+ \tilde{R}_{\mu\nu}f_R - \frac{g_{\mu\nu}}{2} f = \kappa^2 T_{\mu\nu} - \pc{T_{\mu\nu} + \Theta_{\mu\nu}} f_T, 
\end{equation}
where $f_R \equiv \frac{\partial f}{\partial R}$, $f_T \equiv \frac{\partial f}{\partial T}$,  $T_{\mu\nu}$ is the energy-momentum tensor that is defined as
\begin{equation}\label{tensor energia}
    T_{\mu\nu} = \frac{-2}{\sqrt{-g}} \frac{\partial \pc{\sqrt{-g}\mathcal{L}_m}}{\partial g^{\mu\nu}} = -2 \frac{\partial \mathcal{L}_m}{\partial g^{\mu\nu}} + g_{\mu\nu} \mathcal{L}_m,
\end{equation}
and the tensor $\Theta_{\mu\nu}$ describes the variation of the energy-momentum tensor with respect to the metric 
\begin{equation}
    \Theta_{\mu\nu} \equiv \frac{\delta T_{\alpha\beta}}{\delta g^{\mu\nu}} g^{\alpha\beta}.
\end{equation}
Using eq.(\ref{tensor energia}) this tensor is written as
\begin{equation}
    \Theta_{\mu\nu} = -2T_{\mu\nu} + g_{\mu\nu} \mathcal{L}_m  -2 g^{\alpha\beta} \frac{\partial^2 \mathcal{L}_m}{\partial g^{\mu\nu}\partial g^{\alpha\beta}}.\label{ThetaM}
\end{equation}

Contracting eq.(\ref{field_1}) with the metric, the Ricci-Palatini scalar is given as
\begin{equation}\label{field_2}
    R\left(g,\tilde{\Gamma}\right) f_R = \kappa^2 T - \pc {T + \Theta} f_T + 2f + R,
\end{equation}
where $\Theta=\Theta^\mu\,_\mu$. Using the Einstein-Palatini tensor
\begin{equation}
    G_{\mu\nu} \pc{g, \tilde{\Gamma}} = \tilde{R}_{\mu\nu} \pc{\tilde{\Gamma}} - \frac{1}{2} g_{\mu\nu} \tilde{R} \pc{g,\tilde{\Gamma}}, 
\end{equation}
and eqs.(\ref{field_1}) and (\ref{field_2}), the field equation from metric variation becomes
\bea \label{conteudo_materia}
     G_{\mu\nu} \pc{g, \tilde{\Gamma}} &=& \frac{1}{f_R} \{ \kappa^2 T_{\mu\nu} - f_T\pc{T_{\mu\nu} + \Theta_{\mu\nu}}-G_{\mu\nu}(g)
      - \frac{g_{\mu\nu}}{2} \pr{f+\kappa^2 T - f_T\pc{ T+\Theta} + R}\}.
\eea
As the action depends on both the metric and the connection,  now the variation of the action is performed in relation to the $\tilde{\Gamma}$, keeping the metric constant. Then
\begin{align}
    \delta f &= f_R \delta R\pc{\tilde{\Gamma}}, \\
    &= f_R g^{\mu\nu} \delta \tilde{R}_{\mu\nu} \pc{\tilde{\Gamma}}, \\
    &= f_R g^{\mu\nu} \pr{\tilde{\nabla} \pc{\delta \tilde{\Gamma}_{\mu\nu}^{\lambda}} - \tilde{\nabla} \pc{\delta \tilde{\Gamma}_{\mu\lambda}^{\lambda}}},
\end{align}
where $\tilde{\nabla}$ is the covariant derivative associated with $\tilde{\Gamma}$. Then the variation of the action becomes
\begin{equation}
    \delta S = \frac{1}{\kappa ^2} \int \sqrt{-g} f_R g^{\mu\nu} \pr{\tilde{\nabla} \pc{\delta \tilde{\Gamma}_{\mu\nu}^{\lambda}} - \tilde{\nabla} \pc{\delta \tilde{\Gamma}_{\mu\lambda}^{\lambda}}} d^4x.
\end{equation}
Defining $A^{\mu\nu} \equiv f_Rg^{\mu\nu}$ and integrating by parts, we get
\bea
    \kappa^2 \delta S &=& \int \tilde{\nabla}_\lambda \pr{\sqrt{-g} \pc{A^{\mu\nu} \delta \tilde{\Gamma}_{\mu\nu}^{\lambda} -A^{\mu\lambda} \delta \tilde{\Gamma}_{\mu\alpha}^{\alpha}}}d^4x
-    \int \tilde{\nabla}_\lambda \pr{\sqrt{-g} \pc{A^{\mu\nu} \delta^\lambda_\alpha - A^{\mu\lambda} \delta^\nu_\alpha}} \delta \tilde{\Gamma}_{\mu\nu}^{\alpha} d^4x.
\eea
Note that the first term is a total derivative that can be related to a surface integral that becomes zero. This leads to
\begin{equation}\label{integral}
    \tilde{\nabla}_\lambda \pr{ \sqrt{-g} \pc { A^{\mu\nu} \delta^\lambda_\alpha - A^{\mu\lambda} \delta^\nu_\alpha}} = 0.  
\end{equation}
The case $\lambda \neq \alpha$ implies that
\begin{equation}\label{conformal_1}
    \tilde{\nabla}_\lambda \pr{ \sqrt{-g} f_R g^{\mu\nu}} =0
\end{equation}
while for case $\lambda = \alpha$, eq. (\ref{integral}) is identically zero.

The connection $\tilde{\Gamma}$ is compatible with the conformal metric $\tilde{g}_{\mu\nu}$, which means that the geometry is not modified. Then the Palatini connection is written as
\begin{equation}
    \tilde{\Gamma}_{\mu\nu}^{\lambda} = \frac{1}{2} \tilde{g}^{\lambda\rho} \pc{\partial_\nu\tilde{g}_{\rho\mu} + \partial_\mu\tilde{g}_{\rho\nu} + \partial_\rho\tilde{g}_{\mu\nu}}.
\end{equation}

In terms of the Levi-Civita connection $\Gamma$, the components of $\tilde{\Gamma}$ are
\begin{equation}
    \tilde{\Gamma}_{\mu\nu}^{\lambda} = \Gamma_{\mu\nu}^{\lambda} + \frac{1}{2f_R} \pc{ \delta^\lambda_\mu \partial_\nu + \delta^\lambda_\nu \partial_\mu - g_{\mu\nu} \partial^\lambda } f_R.
\end{equation}

Using this conformal transformation and the properties of a conformal metric, the Ricci tensor becomes 
\bea
    \tilde{R}_{\mu\nu}&=&R_{\mu\nu} + \frac{1}{f_R} \Bigl\{\frac{3}{2F_r} \nabla_\mu f_R \nabla_\nu f_R
     - \pc{\nabla_\mu \nabla_\nu + \frac{g_{\mu\nu}}{2} \Box } f_R\Bigl\}.
\eea
The conformal Ricci scalar is given by $\tilde{R} \equiv \tilde{g}^{\mu\nu} \tilde{R}_{\mu\nu} = f_R^{-1} R$ and $\tilde{G}_{\mu\nu} \equiv \tilde{R}_{\mu\nu} - \tilde{g}_{\mu\nu}\tilde{R} /2= G_{\mu\nu} \pc{g, \tilde{\Gamma}}$. Remember that the quantities with tilde are related to the conformal metric. Then the Ricci scalar is written as
\begin{equation}
    \tilde{R} = R - \frac{3}{f_R} \Box f_R + \frac{3}{2f_R^2} \pc{\nabla f_R}^2,
\end{equation}
and the Einstein tensor is 
\bea
    \tilde{G}_{\mu\nu} &=& G_{\mu\nu} + \frac{1}{f_R} \Bigl\{ \pc{g_{\mu\nu} \Box - \nabla_\mu \nabla_\nu} f_R
     + \frac{3}{2f_R} \pr{\nabla_\mu f_R \nabla_\nu f_R - \frac{g_{\mu\nu}}{2} \pc{\nabla f_R}^2 }\Bigl\}.  
\eea

Finally, from eq. (\ref{conteudo_materia}), the field equation for this theory becomes
\bea\label{eq_campo1}
    \pr{1 + f_R} G_{\mu\nu} &+& J_{\mu\nu} = \kappa^2 T_{\mu\nu} - f_T\pc{T_{\mu\nu} + \Theta_{\mu\nu}} 
     - \frac{g_{\mu\nu}}{2} \pr{f + \kappa^2 T - f_T \pc{T + \Theta } + R } ,
\eea
where, for convenience, it is defined
\bea
    J_{\mu\nu} &\equiv&  \pc{g_{\mu\nu} \Box - \nabla_\mu \nabla_\nu} f_R
     + \frac{3}{2f_R} \pr{\nabla_\mu f_R \nabla_\nu f_R - \frac{g_{\mu\nu}}{2} \pc{\nabla f_R}^2 }.
\eea

In the next section, the field equations of the hybrid metric-Palatini $f(\R,T)$ gravity are solved for the G\"{o}del-type metrics.


\section{Gödel-type Metric in Hybrid Metric-Palatini $f(\R,T)$ gravity}
In this section, the main features of a generalized class of metrics called G\"{o}del-type are briefly discussed. Then, the question of causality in the hybrid metric-Palatini $f(\R,T)$ gravity is investigated. The line element that describes the G\"{o}del-type metric in cylindrical coordinates takes the following form
\begin{equation}\label{type_godel}
    ds^2 = -dt^2 - 2H(r) dtd\phi + dr^2 + G(r) d\phi^2 + dz^2,
\end{equation}
where $G(r) = D^2(r) - H^2(r)$ and the functions $D(r)$ and $H(r)$ satisfy the relations
\bea
\frac{H'(r)}{D(r)}&=&2\omega,\nonumber\\
\frac{D''(r)}{D(r)}&=&m^2,
\eea
with the prime being the derivative with respect to the radius coordinate. It is important to note that, the parameters $\omega$ and $m$ describe entirely the conditions of homogeneity of space-time. These parameters assume the values in the range $-\infty \le m^2  \le +\infty$ and $\omega\neq 0$. As shown in \cite{tipo_godel1}, G\"{o}del-type metrics can be divided into three different classes: (i) hyperbolic class, where $m^2>0$ and $\omega\neq 0$; (ii) trigonometric class, with $m^2<0$ and $\omega\neq 0$ and (iii) linear class, where $m^2=0$ and $\omega\neq 0$. The G\"{o}del metric, which is obtained for the case $m^2=2\omega^2$, belongs to the hyperbolic class.

The main characteristic of G\"{o}del-type spaces is the possibility of CTCs that are cicles defined by $t,r,z=\mathrm{const}$ and $\phi\in (0,2\pi)$ in a region limited by the range $r_1<r<r_2$, with $G(r)$ being negative within this range. For the hyperbolic class, the functions $H(r)$ and $D(r)$ are given as
\bea
H(r)&=&\frac{4\omega}{m^2}sinh^2\left(\frac{mr}{2}\right),\\
D(r)&=&\frac{1}{m}sinh(mr),
\eea
and CTCs arise within the region corresponding to $r > r_c$, where $r_c$ is the critical radius separating the causal and non-causal regions. The explicit form of the critical radius is given by
\bea
sinh^2\left(\frac{mr_c}{2}\right)=\left(\frac{4\omega^2}{m^2}-1\right)^{-1}.
\eea
It is interesting to note two special cases: (i) for $m^2=2\omega^2$ the standard G\"{o}del solution is recovered and its critical radius becomes
\begin{equation}\label{eq:112}
r_c=\frac{2}{m}sinh^{-1}(1);
\end{equation}
and (ii) for $m^2=4\omega^2$ there are no CTCs, since the critical radius $r_c = \infty$, that is, the violation of causality is avoided. In addition, similarly, linear and trigonometric classes also display CTCs.

Therefore, in this theory, there is the possibility of a causal rather than a non-causal solution. These regions are determined from the free metric parameters, i.e., $m$ and $\omega$, and limited by a critical radius $r_c$. Then, the G\"{o}del-type solutions bring more details to the problem of causality. 

In order to make calculations simpler, let's define a local set of tetrad basis $\theta^A=e^A\,_\mu dx^\mu$, so that the line element takes the form
\begin{equation}
    ds^2 = \eta_{AB} \theta^A \theta^B = (\theta^0)^2 - (\theta^1)^2 - (\theta^2)^2 - (\theta^3)^2, \label{frame}
\end{equation}
where the capital Latin letters denote the transformed space, $\eta_{AB}$ is the Minkowski metric and
\begin{align}
    \theta^{(0)} &= dt + H(r)d\phi, \\ \theta^{(1)} &= dr, \\ \theta^{(2)} &= D(r)d\phi, \\ \theta^{(3)} &= dz.    
\end{align}

The non-zero components of the Einstein tensor in the flat (local) space-time take the form,
\begin{align}
     G_{(0)(0)} &= 3\omega^2-m^2, \\
    G_{(1)(1)} &= \omega^2,\\
    G_{(2)(2)} &= \omega^2, \\
    G_{(3)(3)} &= m^2-\omega^2,
\end{align}
where $G_{AB}=e^\mu_A e^\nu_B G_{\mu\nu}$ has been used. Here, $e^\mu\,_B$ is the inverse of $e^A\,_\mu$ and they satisfy the relation  $ e^A\,_\mu e^\mu\,_B=\delta^A_B$. The non-zero components of the tetrads are
\begin{equation}
        e^{(0)}\ _{0} = e^{(1)}\ _{1} = e^{(3)}\ _{3} = 1, e^{(0)}\ _{2} = H(r),  e^{(2)}\ _{2} = D(r),
\end{equation}
and the non-zero components of the inverse of the tetrad are
\bea
        e^{0}\ _{(0)} &=& e^{1}\ _{(1)} = e^{3}\ _{(3)} = 1,\nonumber\\
         e^{0}\ _{(2)} &=& - \frac{H(r)}{D(r)}, e^{2}\ _{(2)} = D^{-1}(r).
\eea

Thus, the field equation (\ref{eq_campo1}) in the tetrad basis becomes
\bea\label{eq_campo_tipo_godel}
    \pr{1+f_R} G_{AB} &+& J_{AB} = \kappa^2 T_{AB} - f_T\pc{T_{AB} + \Theta_{AB}}
     - \frac{g_{AB}}{2} \pr{f + \kappa^2 T - f_T \pc{T + \Theta} + R }. 
\eea
To complete the field equations, it is necessary to fix the matter content. In \cite{tipo_godel1} well-motivated matter sources such as a perfect fluid, a scalar field, and an electromagnetic field are investigated. Here, a perfect fluid, a scalar field, and a combination of both are considered.

\subsection{Perfect Fluid}

Let's start by assuming that the content of matter is a perfect fluid whose energy-momentum tensor in the local frame is given as
\begin{equation}
    T_{AB}=(\rho+p)u_A u_B-p\eta_{AB},
\end{equation}
where $u_A = (1,0,0,0)$ is the four-velocity of the fluid and $\rho$ and $p$ are the fluid density and pressure, respectively. It is important to note that, the Lagrangian of matter that describes a perfect fluid in gravitational models with non-minimal coupling between matter and geometry has been intensely explored \cite{Faraoni, Bertolami, Bertolami1}. These theories carry an arbitrariness, that is, the Lagrangian of matter is not uniquely defined. However, there are different studies in which it is shown that such arbitrariness can be avoided \cite{Harko10, Moraes}.

For this matter source, the field equations are given as
{\small
\begin{align}
    2\pr{1+f_R}f_R\pc{3\omega^2 - m^2} +f  &= 8\pi \pc {\rho + 3p} + \pc{\rho +p}f_T + R, \\
    2\pr{1+f_R}f_R\omega^2 - f &= 8\pi \pc {\rho - p} + \pc{\rho +p}f_T - R, \label{resolvendo_tipo1}\\ 
    2\pr{1+f_R}f_R\pc{m^2 - \omega^2} -f &= 8\pi \pc {\rho - p} + \pc{\rho +p}f_T - R, \label{resolvendo_tipo2}
\end{align}}
where the Ricci scalar for the G\"{o}del-type metrics takes a constant value $R = 2(m^2 - \omega^2)$. From eq.(\ref{resolvendo_tipo1}) and eq.(\ref{resolvendo_tipo2}) is found that
\begin{equation}
    2\pc{1+f_R}(2\omega^2-m^2)f_R=0.\label{G}
\end{equation}
Note that, for $f_R>0$, eq. (\ref{G}) leads to $m^2=2\omega^2$ which defines the G\"{o}del metric and the remaining field equations provide
\bea
\rho&=&\frac{-f+m^2}{16\pi}+\frac{(1+f_R)f_R}{8\pi+f_T}m^2,\\
p&=&\frac{f-m^2}{16\pi}.
\eea
Analyzing these solutions, it is observed that the G\"{o}del solution is allowed. Then, inevitably closed timelike curves are exhibited, i.e. there are non-causal G\"{o}del's circles which are limited by the critical radius $r_c$. In the framework of hybrid metric-Palatini $f(\R,T)$ gravity, $r_c$ is given as
\bea
r_c=\frac{2}{m}\sinh^{-1}(1)=2\sinh^{-1}(1)\sqrt{\frac{(1+f_R)f_R}{(\rho+p)(f_T+8\pi)}}.
\eea
This result can be divided as follows. First, the G\"{o}del-type metric is a solution of the hybrid metric-Palatini $f(\R,T)$ gravity. Second, the critical radius depends on both gravity theory and matter content. Therefore, the main conclusion is that the violation of causality is allowed in this gravitational theory. Now let's investigate whether other sources of matter can generate causal solutions.

\subsection{Perfect fluid and scalar field}

In this section, in order to find a causal G\"{o}del-type solution, a combined energy-momentum tensor is considered, that is,
\bea
    T_{AB} &=& T_{AB}^m + T_{AB}^S,\nonumber \\ 
    &=& \pc{\rho + p} u_Au_B -p\eta_{AB} + \nabla_A \phi \nabla_B \phi
     - \frac{1}{2} \eta_{AB}\eta^{CD}\nabla_C \phi \nabla_D \phi,
\eea
where $T_{AB}^m$ and $T_{AB}^S$ are the energy-momentum tensor to a perfect fluid and a scalar field, respectively. Here, $\nabla_A$ is the covariant derivative relative to the local basis $\theta^A = e^A_\beta dx^\beta$ and the scalar field is chosen as $\phi(z) = \epsilon z + \epsilon $, with $\epsilon = \mathrm{const}$. The non-zero components of the energy-momentum tensor associated with the scalar field are,
\begin{equation}
    T_{00}^S = -T_{11}^S = -T_{22}^S = T_{33}^S = \frac{\epsilon^2}{2},
\end{equation}
and its trace is given by
\begin{equation}
    T^S=\rho -3p+\epsilon^2.
\end{equation}

Considering the contributions of the scalar field, the tensor $\Theta_{AB}$ is rewritten  as
\begin{equation}
    \Theta_{AB} = \Theta_{AB}^m + \Theta_{AB}^S,
\end{equation}
where $\Theta_{AB}^m$ is associated with the perfect fluid and is given as
\bea
\Theta_{AB}^m=-2T_{AB}-pg_{AB}.
\eea
Here, eq. (\ref{ThetaM}) is used and the matter Lagrangian is taken as ${\cal L}_m=-p.$ To calculate $\Theta_{AB}^S$, let's consider the following Lagrangian,
\begin{equation}
    \mathcal{L}^S = \eta^{AB} \nabla_A \phi \nabla_B \phi, 
\end{equation}
such that the tensor becomes,
\begin{equation}
    \Theta_{AB}^S = - T_{AB}^S + \frac{1}{2} T^S \eta_{AB}.
\end{equation}

Then the field equation (\ref{eq_campo_tipo_godel}) is written as
\bea
    \pr{1+f_R} G_{AB} &=& \kappa^2 \pr {\pc{\rho  + p}u_Au_B -p\eta_{AB} + T_{AB}^S }
    -\frac{1}{2} \Bigl\{\kappa^2 \pc{\rho - 3p + \epsilon^2 } +f\nonumber\\
     &+&f_T \pc{\rho + p - 2\epsilon^2 } + R \Bigl\}\eta_{AB}
    +f_T \pr{\pc{\rho + p }u_Au_B - \frac{1}{2} \epsilon^2 \eta_{AB}}.
\eea

Using the G\"{o}del-type metric, a set of equations is obtained, i.e.,
{\small
\begin{align}
    \kappa^2 \epsilon^2 &= \pc{m^2 - 2\omega^2} \pr{1+f_R} + R,\\
    \kappa^2 p + \frac{1}{2}\epsilon^2 f_T &= \frac{1}{2} \pc{2\omega^2 - m^2}\pr{1+f_R} +\frac{1}{2} f - R,\\
    \kappa^2 \rho + f_T \pc{ \rho + p -\frac{1}{2}\epsilon^2} &= \frac{1}{2} \pc{6\omega^2 - m^2} \pr{1+f_R} -\frac{1}{2}f -R.
\end{align}}
Assuming that $f_R > 0$ and $f_T > 0$, these equations allow for a causal G\"{o}del-type class of solutions that is given as
\bea
    m^2 &=& 4\omega^2,\label{eq65}\\
    f_R &=& \frac{\kappa^2\epsilon^2 - 6\omega^2}{2\omega^2} - 1,\\
    f_T&=&-\frac{1}{\rho+p}[\kappa^2(\rho+p)-12\omega^2].
\eea
Note that, eq. (\ref{eq65}) leads to $r_c \rightarrow \infty$.  Therefore, for matter content, which is a combination of a perfect fluid and a scalar field, causality is not violated for any hybrid metric-Palatini $f(\R,T)$ gravity.

\subsection{Scalar field}

If only the scalar field $\phi(z)$ is considered, the field equations become
\begin{align}
    \pr{1+f_R}\pc{3\omega^2 - m^2} + \frac{f}{2} &= \frac{1}{2}\epsilon^2 f_T + R,\\
    \pr{1+f_R}\omega^2 - \frac{f}{2} &= - \frac{1}{2} \epsilon^2 f_T - R,\\
    \pr{1+f_R}\pc{m^2 - \omega^2} -\frac{f}{2} &= 8\pi \epsilon^2 -\frac{1}{2} f_T \epsilon^2 - R.
\end{align}
These equations allow the solution
\bea
m^2&=&4\omega^2.
\eea
Therefore, when only the scalar field is considered, this theory defines a class of solutions without causality violation, since the critical radius goes to infinity. Furthermore, this result holds for an arbitrary $f(\R,T)$ function with $f_R\neq 0$ and $f_T\neq 0$.

\section{Discussions and Final Remarks}

In this paper, the hybrid metric-Palatini $f(\R, T)$ gravity is considered. This modified theory of gravity consists of adding to the Einstein–Hilbert Lagrangian a function $f(\R,T)$ that is constructed in the Palatini formalism. Thus, this theory combines the metric and Palatini approaches to gravity and is an extension of $f(R,T)$ theories, which define a non-minimal coupling between matter and geometry. Then the modified Einstein field equations are obtained. It is important to note that if this theory describes gravity, there are several issues that must be investigated in this context. A question that must be re-examined is whether this theory of gravity allows for G\"{o}del-type solutions, which lead to the violation of causality. Although locally the general relativity has the same causal structure as the flat space-time of special relativity, on a non-local scale it permits solutions to Einstein field equations that show causal anomalies such as CTC's. Here, the question arises: is the causality violation allowed in the hybrid metric-Palatini  $f(\R, T)$ gravity? In order to answer this question, a combination of physical well-motivated sources of matter is considered. For the perfect fluid as the content of matter is shown that the violation of causality is an unavoidable characteristic of any hybrid metric-Palatini  $f(\R, T)$ gravity. Then, an expression for the critical radius has been derived. It depends on both the gravity theory and the content of matter. To find a causal G\"{o}del-type solution, two different types of matter have been introduced: (i) a combination of perfect fluid and scalar field, and (ii) just a scalar field. In both cases, causal solutions are permitted, that is, there is a condition between the metric parameters $m$ and $\omega$ such that the critical radius becomes infinity. Therefore, the existence of causal and non-causal G\"{o}del-type solutions makes it clear that hybrid metric-Palatini $f(\R, T)$ gravity does not prevent causal anomalies in the form of CTCs that are allowed in Einstein theory of gravity. It is important to note that in \cite{Azizi} the G\"{o}del-type solution and the causality violation is studied in Palatini $f(R)$ gravity with a non-minimal curvature-matter coupling. However, this model is a type of $f(R, {\cal L}_m)$ gravity. Here, our model is different, it is an extension of $f(R,T)$ in the hybrid formalism, while in this reference only the Palatini formalism is considered. Therefore the theories are different and the conditions required  for a causal and non-causal solution are different.

\section*{Acknowledgments}
This work by A. F. S. is supported by CNPq projects 430194/2018-8 and 313400/2020-2; J. S. Gon\c{c}alves thanks CAPES for financial support.

\end{document}